\documentclass[twocolumn, times]{aastex6}
\let\pwifjournal=\iffalse \def\version{Manuscript of 2016/08/15}

\usepackage{amsmath}

\usepackage[T1]{fontenc}
\pwifjournal\else
  \usepackage{microtype}
\fi

\pwifjournal\else
  \makeatletter
  \renewcommand\plotone[1]{%
    \centering \leavevmode \setlength{\plot@width}{0.99\linewidth}
    \includegraphics[width={\eps@scaling\plot@width}]{#1}%
  }%
  \makeatother
\fi

\pwifjournal\else
  \usepackage{etoolbox}
  \makeatletter
  \patchcmd{\NAT@citex}
    {\@citea\NAT@hyper@{%
       \NAT@nmfmt{\NAT@nm}%
       \hyper@natlinkbreak{\NAT@aysep\NAT@spacechar}{\@citeb\@extra@b@citeb}%
       \NAT@date}}
    {\@citea\NAT@nmfmt{\NAT@nm}%
     \NAT@aysep\NAT@spacechar\NAT@hyper@{\NAT@date}}{}{}
  \patchcmd{\NAT@citex}
    {\@citea\NAT@hyper@{%
       \NAT@nmfmt{\NAT@nm}%
       \hyper@natlinkbreak{\NAT@spacechar\NAT@@open\if*#1*\else#1\NAT@spacechar\fi}%
         {\@citeb\@extra@b@citeb}%
       \NAT@date}}
    {\@citea\NAT@nmfmt{\NAT@nm}%
     \NAT@spacechar\NAT@@open\if*#1*\else#1\NAT@spacechar\fi\NAT@hyper@{\NAT@date}}
    {}{}
  \makeatother
\fi

\usepackage{natbib}
\makeatletter
\newcommand\pkgw@simpfx{http://simbad.u-strasbg.fr/simbad/sim-id?Ident=}
\newcommand\MakeObj[4][\@empty]{
  \pwifjournal%
    \expandafter\newcommand\csname pkgwobj@c@#2\endcsname[1]{\protect\object[#4]{##1}}%
  \else%
    \expandafter\newcommand\csname pkgwobj@c@#2\endcsname[1]{\href{\pkgw@simpfx #3}{##1}}%
  \fi%
  \expandafter\newcommand\csname pkgwobj@f#2\endcsname{#4}%
  \ifx\@empty#1%
    \expandafter\newcommand\csname pkgwobj@s#2\endcsname{#4}%
  \else%
    \expandafter\newcommand\csname pkgwobj@s#2\endcsname{#1}%
  \fi}%
\newcommand{\obj}[1]{%
  \expandafter\ifx\csname pkgwobj@c@#1\endcsname\relax%
    \textbf{[unknown object!]}%
  \else%
    \csname pkgwobj@c@#1\endcsname{\csname pkgwobj@s#1\endcsname}%
  \fi}
\newcommand{\objf}[1]{%
  \expandafter\ifx\csname pkgwobj@c@#1\endcsname\relax%
    \textbf{[unknown object!]}%
  \else%
    \csname pkgwobj@c@#1\endcsname{\csname pkgwobj@f#1\endcsname}%
  \fi}
\makeatother

\MakeObj[WISE~1122$+$25]{1122+25}{WISEP\%20J112254.73\%2b255021.5}{WISEP~J112254.73$+$255021.5}
\MakeObj{phasecal}{NVSS\%20J112555\%2b260630}{NVSS~J112555$+$260630}
\MakeObj{3c286}{3C\%20286}{3C~286}
\MakeObj{lhs302}{LHS\%20302}{LHS~302}
\MakeObj[2M~1047$+$21]{1047+21}{2MASSI\%20J1047539\%2b212423}{2MASSI~J1047539$+$212423}
\MakeObj{n33370}{NLTT\%2033370}{NLTT~33370~AB}
\MakeObj{1048-39}{DENIS\%20J1048.0-3956}{DENIS~J1048.0$-$3956}
\MakeObj{1237+65}{2MASS\%20J12373919\%2b6526148}{2MASS~J12373919$+$6526148}

\newcommand\apx{\ensuremath{\sim}}
\newcommand\citeeg[1]{\citep[\emph{e.g.},][]{#1}}
\renewcommand\deg{\ensuremath{^\circ}}
\newcommand\dt{\ensuremath{{\Delta t}}}

\newcommand\ha{\ensuremath{\text{H}\alpha}}

\newcommand\kms{km~s$^{-1}$}
\newcommand\lbol{\ensuremath{\text{L}_\text{bol}}}
\newcommand\lnura{\ensuremath{\langle\text{L}_{\nu,\text{R}}\rangle}}
\newcommand\lsun{\ensuremath{\text{L}_\odot}}
\newcommand\rjup{\ensuremath{\text{R}_\text{J}}}

\newcommand\ujy{$\mu$Jy}
\newcommand\ujybm{$\mu$Jy~bm$^{-1}$}

\begin{document}

\title{Variable and Polarized Radio Emission from the T6 Brown Dwarf WISEP~J112254.73$+$255021.5}
\author{
  P.~K.~G. Williams\altaffilmark{1},
  J.~E. Gizis\altaffilmark{2},
  E. Berger\altaffilmark{1}
}

\altaffiltext{1}{Harvard-Smithsonian Center for Astrophysics, 60 Garden Street,
  Cambridge, MA 02138, USA}
\altaffiltext{2}{Department of Physics and Astronomy, University of Delaware, Newark, DE
  19716, USA}
\email{pwilliams@cfa.harvard.edu}

\slugcomment{\version}
\shorttitle{Variable and Polarized Radio Emission from T6 Brown Dwarf}
\shortauthors{Williams, Gizis, Berger}

\begin{abstract}
  Route \& Wolszczan (2016) recently detected five radio bursts from the
  T6~dwarf WISEP~J112254.73$+$255021.5 and used the timing of these events to
  propose that this object rotates with an ultra-short period of
  \apx17.3~minutes. We conducted follow-up observations with the Very Large
  Array and Gemini-North but found no evidence for this periodicity. We do,
  however, observe variable, highly circularly polarized radio emission
  possibly with a period of 116~minutes, although our observation lasted only
  162~minutes and so more data are needed to confirm it. Our proposed
  periodicity is typical of other radio-active ultracool dwarfs. The
  handedness of the circular polarization alternates with time and there is no
  evidence for any unpolarized emission component, the first time such a
  phenomenology has been observed in radio studies of very low-mass stars and
  brown dwarfs. We suggest that the object's magnetic dipole axis may be
  highly misaligned relative to its rotation axis.
\end{abstract}


\section{Introduction}

It has long been recognized that some M~dwarf stars vary both
spectroscopically and photometrically \citep{l26, vm40}. Early on, theorists
concluded that despite their small sizes, these stars must harbor magnetic
field more powerful than those of the Sun \citeeg{s67}, although this
hypothesis was not confirmed until the pioneering observations of
\citet{sl85}. While theoretical considerations drove the assumption that the
coolest M~dwarfs would not be able to generate magnetic fields as strong as
those of the classical flare stars \citeeg{ddyr93}, radio observations have
demonstrated that strong dynamo action extends not only to more massive brown
dwarfs \citep{bbb+01} but even to the T~dwarfs, which have effective
temperatures of just $\approx$1000~K \citep{rw12, rw16, khp+16}. Evidently the
loss of the ``tachocline'' --- the shearing layer between the radiative core
and convective outer envelope that is argued to be important to the solar
dynamo \citep[and references therein]{o03} --- does not preclude the
generation of strong magnetic fields in these objects, which are fully
convective.

There is nonetheless ample evidence that the magnetic properties of the
ultracool dwarfs --- very low mass stars and brown dwarfs with spectral types
M7 or later \citep{krl+99, mdb+99} --- are substantially different than those
of their higher-mass brethren. Their X-ray and \ha\ emission drop
precipitously with temperature \citep{gmr+00, whw+04, smf+06, bbf+10}, so that
ratios of their radio and X-ray luminosities significantly exceed standard
relations for active stars \citep{gb93, bbb+01, wcb14}; a new phenomenology of
periodic, highly polarized radio bursts arises \citep{brr+05, brpb+09, had+06,
  had+08}; trends relating rotation and X-ray activity break down
\citep{bbg+08, cwb14}; and their activity lifetimes become much longer than
those of Sun-like stars, implying inefficient angular momentum loss and rapid
rotation even to \apx Gyr ages \citep{grh02, whb+08, bmm+14}.

\citet{rw16} discovered that the T6 \objf{1122+25} (hereafter \obj{1122+25})
is one of the handful of known radio-emitting T~dwarfs \citep{rw12, wbz13,
  khp+16} in the course of a large Arecibo survey for radio bursts from
ultracool dwarfs at 5~GHz \citep{rw13}. Along with being an extreme object in
terms of its temperature, \citet{rw16} claimed that it is extreme in terms of
rotation: they phased the arrival time of its radio bursts
(\autoref{s.target}) to infer a rotation period of \apx17.3~minutes. If
confirmed, this rotation period would be by far the shortest ever measured in
a brown dwarf. \obj{1122+25} may therefore be a unique laboratory for
understanding the relationship between age, rotation, and magnetic field
generation in the fully-convective dynamo regime.

The outline of this paper is as follows. We first summarize previous
observations of \obj{1122+25} (\autoref{s.target}). We then describe new
follow-up observations that we obtained with the Karl G. Jansky Very Large
Array (VLA) and Gemini-North and present the data (\autoref{s.obs}). Next we
analyze the data for periodic signals, finding evidence for variations at a
period of \apx116~min rather than the value suggested by \citet{rw16}, and
analyze the properties of the observed radio emission
(\autoref{s.radioprops}). Finally, we discuss the implications of these
results (\autoref{s.disc}) and present our conclusions (\autoref{s.conc}).

\section{The Radio-Active T6 Dwarf \objf{1122+25}}
\label{s.target}

\citet{kcg+11} identified \obj{1122+25} as a candidate T~dwarf in
\textit{Widefield Infrared Survey Explorer} (WISE) imaging and confirmed its
cool nature spectroscopically, assigning it a NIR spectral type of T6.
Additional analysis has yielded a spectrophotometrically estimated distance of
17.7~pc \citep{kgc+12}.

Over a span of four observing epochs in their 5~GHz Arecibo survey,
\citet{rw13, rw16} detected five radio bursts from \obj{1122+25}. The peak
fluxes were \apx1.5--3~mJy, and two of the bursts were separated by
\apx18.3~minutes. In ten additional, non-contiguous hours of follow-up
observations, additional flares were not detected; the total time on source
was 29~hr. The detected flares had left circular polarizion (LCP) fractions
ranging from 15--100\% and characteristic durations of 30--120~s. \citet{rw16}
did not find a quiescent counterpart to the radio source in the FIRST catalog
of radio sources \citep{bwh95}.

The high circular polarization fraction and brightness temperature of
\obj{1122+25}'s radio bursts are consistent with emission due to the electron
cyclotron maser instability \citep[ECMI;][]{the.ecmi, t06}, as also observed
in a substantial fraction of the radio-active ultracool dwarfs \citep{bp05,
  had+06, had+08, brpb+09, rw12, wb15}. Because this emission is expected to
occur at the local electron cyclotron frequency, $\nu_c = e B / 2\pi m_e c
\simeq 2.8\text{ GHz} (B / \text{kG})$, radio detection provides a measurement
of the magnetic field strength at the site of the radio emission. In the case
of \obj{1122+25}, \citet{rw16} estimated $B \gtrsim 1.8$~kG since the upper
spectral cutoff was not observed.

\citet{rw16} also applied pulsar timing techniques to search for a periodicity
in the times-of-arrival (TOAs) of \obj{1122+25}'s flares. They obtained a
timing solution of $P = 1035.7881 \pm 0.0008\text{ s} \simeq 17.3$~minutes,
with a post-fit rms residual $\sigma \approx 0.86$~s, although they emphasize
that these uncertainties are the formal outputs of the fitting routine and
that the variability in the flare profiles suggests that the true uncertainty
in the period is \apx15~s: a number that is substantially larger but still
small in an absolute sense. The first and last flare detections were separated
by \apx240~days, or \apx20,000 rotations assuming the claimed periodicity.
\citet{rw16} used Monte Carlo simulation to deduce a false-alarm probability
of \apx0.01\% for the period detection. This ultra-short rotation period
implies a highly oblate object rotating near its breakup rate, with an
equatorial velocity $\gtrsim$300~\kms\ \citep{rw16}. The breakup rotation rate
is a function of mean density, which increases with time in the brown dwarf
regime, so that the observed period implies an age $\gtrsim$1~Gyr for
\obj{1122+25} \citep{rw16}. Other constraints on the age of \obj{1122+25} are
not presently available.

The M5 dwarf \obj{lhs302} is found \apx4.4$'$ away from \obj{1122+25} and
appears to have a similar distance and proper motion. If the two objects were
physically related, their projected separation would be \apx4500~AU, so that
interactions between the two would be expected to be negligible
\citep{kcg+11}.

\section{Observations and Data Processing}
\label{s.obs}

\begin{figure*}[tbp]
  \includegraphics[width=\linewidth]{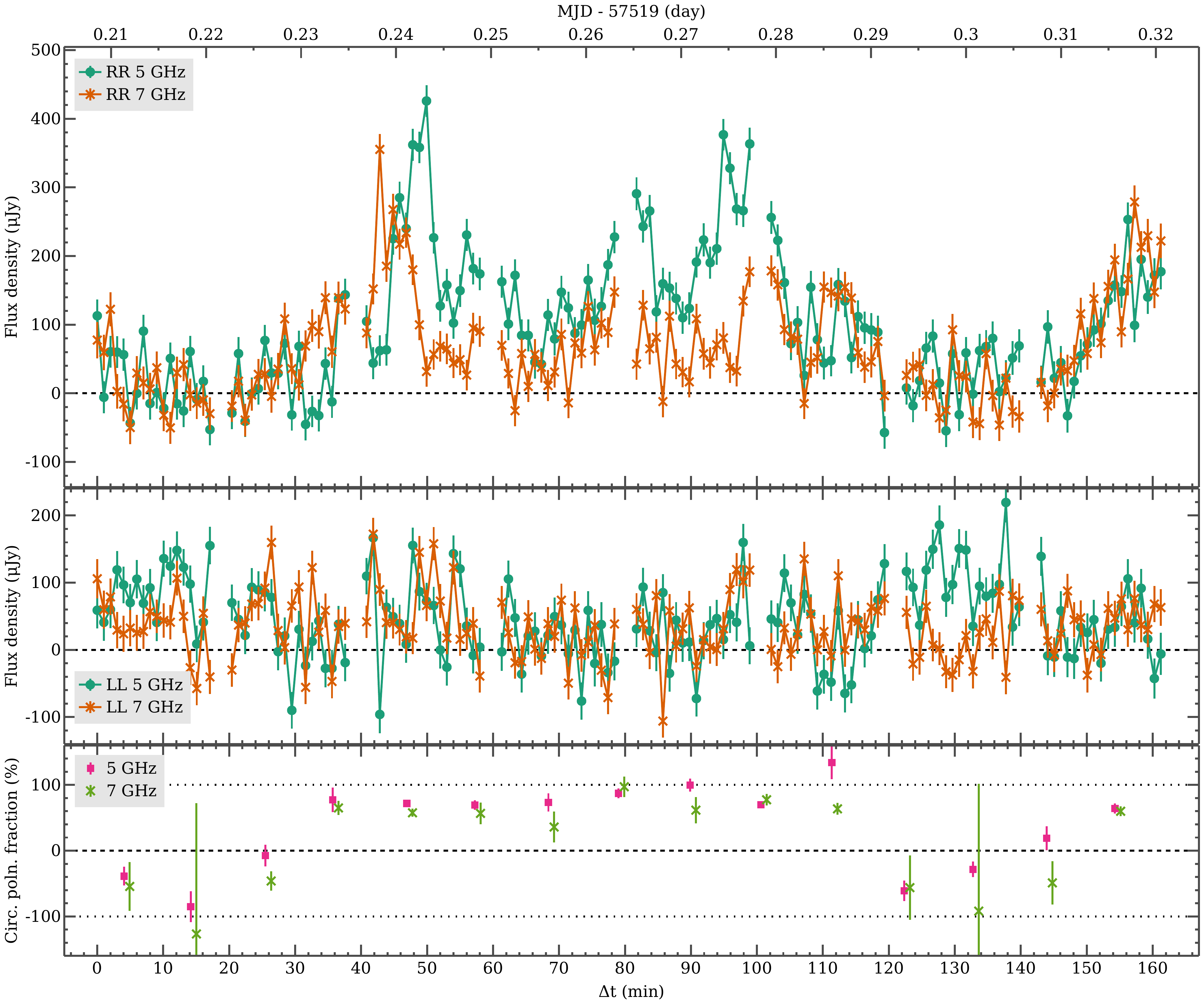}
  \caption{Radio light curves of \obj{1122+25} from the VLA, binned to a
    1-minute cadence from the 10~s cadence at which the visibilities were
    averaged. The total duration of the observations is 162~minutes.
    \textit{Top panel}: light curves in the RR polarization product.
    \textit{middle panel}: light curves in the LL polarization product.
    \textit{Green circles}: data in the 4--6~GHz sub-band. \textit{Orange
      X's}: data in the 6--8~GHz sub-band. Breaks in the lines indicate pauses
    in the observations for calibrator visits. \textit{Bottom panel}: derived
    circular polarization fraction in the 4--6~GHz sub-band, after Hanning
    smoothing the RR and LL data to 10-minute time scales. Full RCP is
    ${+}100$\% and full LCP is ${-}100$\%. Data at the two frequencies are
    slightly offset in time for visual clarity.}
  \label{f.vla}
\end{figure*}

\subsection{Karl G. Jansky Very Large Array}

We obtained Director's Discretionary time to observe \obj{1122+25} for 3~hours
using the VLA on UT date 2016~May~11 (project ID VLA/16A-463; PI: Williams).
The VLA was in the semi-extendend ``CnB'' configuration. We used the VLA's
3-bit samplers to digitize the full bandwidth of the ``C'' band, 4--8~GHz,
dividing the bandpass into channels of 2~MHz width. We obtained standard
calibration observations, using \obj{3c286} as the flux density and bandpass
calibrator and \obj{phasecal} as the complex gain calibrator. The total time
on-source was 2.57~hr.

We used standard procedures to analyze the data, using a Python-based
reduction process driving tasks from the CASA package \citep{the.casa}. We
flagged radio-frequency interference (RFI) automatically using the
morphological and \textsf{SumThreshold} algorithms as implemented by the
\textsf{aoflagger} tool \citep{odbb+10, ovdgr12}. After calibration, we imaged
the total intensity (Stokes~I polarization component) of the full data set at
0.3$\times$0.3 arcsec$^2$ resolution using multifrequency synthesis
\citep{the.mfs} and $w$-projection \citep{cgb05}. The mean frequency of this
image is 6.00~GHz and the rms noise around its phase center is 2.2~\ujybm.

The image contains an isolated source at position RA = 11:22:54.26, Decl. =
$+$25:50:20.0, with uncertainties of 0.2$''$ and 0.1$''$ in RA and Decl.,
respectively. From the astrometric parameters reported by \citet{kcg+11} and
the mean image epoch of MJD~57519.27, the predicted position of \obj{1122+25}
at the time of our observation is RA = 11:22:54.30, Decl. = $+$25:50:20.4,
with uncertainties of 0.5$''$ and 0.6$''$ in RA and Decl., respectively. These
two positions differ by 0.7$''$ or 1.3$\sigma$. Based on this agreement and
the time-variable polarized emission of this source (described below), which
is characteristic of radio-active ultracool dwarfs, we identify this VLA
source with \obj{1122+25}.

We generated two additional images by splitting the 4~GHz of VLA bandwidth
into halves by frequency, resulting in two images with center frequencies of
5.00 and 7.00~GHz. In the low-band image, \obj{1122+25} is a point source with
a time- and polarization- averaged flux density of $77 \pm 4$~\ujy. In the
high-band image, it is a point source with an averaged flux density of $47 \pm
4$~\ujy. The implied spectral index, averaging over time and polarization, is
$\alpha = -1.5 \pm 0.3$, where $S_\nu \propto \nu^\alpha$.

We investigated the time variability of \obj{1122+25} by subtracting the
emission from the other sources in the VLA field of view and summing the
calibrated visibilities after phasing to the source position, as per the
procedure described in \citet{wbz13}. Light curves derived from this analysis
are shown in \autoref{f.vla}, where curves have been binned to a cadence of
60~s (compared to the data sampling time of 5~s) for display purposes. Here
and below we quantify times using \dt, the number of minutes since the
observation start; $\dt = 0$ is $\text{MJD} = 57519.208$.

The emission in the RR polarization product (that is, the mean correlation of
pairs of RCP-sensitive receivers) is highly variable, with several bursts,
predominantly in the 4--6~GHz sub-band (denoted ``5~GHz'' for brevity), having
durations of \apx10~min and peak flux densities of \apx300--400~\ujy. The
second and third 5~GHz RR bursts are separated by \apx16~minutes, close to the
\apx18.3~minute spacing between the two closest Arecibo bursts. This
separation in turn presumably drove the identification of a \apx17.3-minute
periodicity by \citet{rw16}. We note, however, that the Arecibo bursts were
LCP rather than RCP.

The RR emission is generally brighter in the 5~GHz sub-band than the 7~GHz
sub-band, with the exception of the period $30 \lesssim \dt \lesssim 45$,
which ends in a rapid spike. Finer-grained examination of the frequency
dependence of the emission does not reveal any abrupt changes in the spectrum.
If there is a sharp spectral cutoff as expected from the ECMI emission, it
occurs at higher frequencies than we observed.

While the LL polarization product is not as clearly variable, there are two
episodes in which the LL emission at 5~GHz becomes significantly brighter than
that at 6--8~GHz (``7~GHz''): at $\dt \apx 5$ and $\dt \apx 130$. These
episodes coincide with periods of weak or absent emission in the RR
polarization product, so that at most times the radio emission of
\obj{1122+25} is strongly circularly polarized, although the handedness of
that polarization varies. The lower panel of \autoref{f.vla} plots the
polarization fraction as a function of time, smoothing on 10-minute time
scales to increase S/N. The emission is almost always intensely circularly
polarized, with only a few intervals where the absolute value of the
polarization fraction is ${<}50$\%. These intervals seem to correspond to
transitions in the handedness of the polarization, from left to right and back
again.

The Arecibo flares were 15--100\% LCP, had durations of \apx1--3~min, and
reached peak flux densities of 1000-2000~\ujy\ \citep{rw16}. The dominant RR
radio variability that we observe is therefore substantially different in its
gross properties. The episodes of LL emission, however, could represent
higher-frequency and fainter counterparts of the Arecibo flares. Using the
spectral index derived above, the Arecibo flares would have flux densities
\apx1--2~mJy, significantly brighter than the LCP emission we observed.

\begin{figure}[tbp]
  \includegraphics[width=\linewidth]{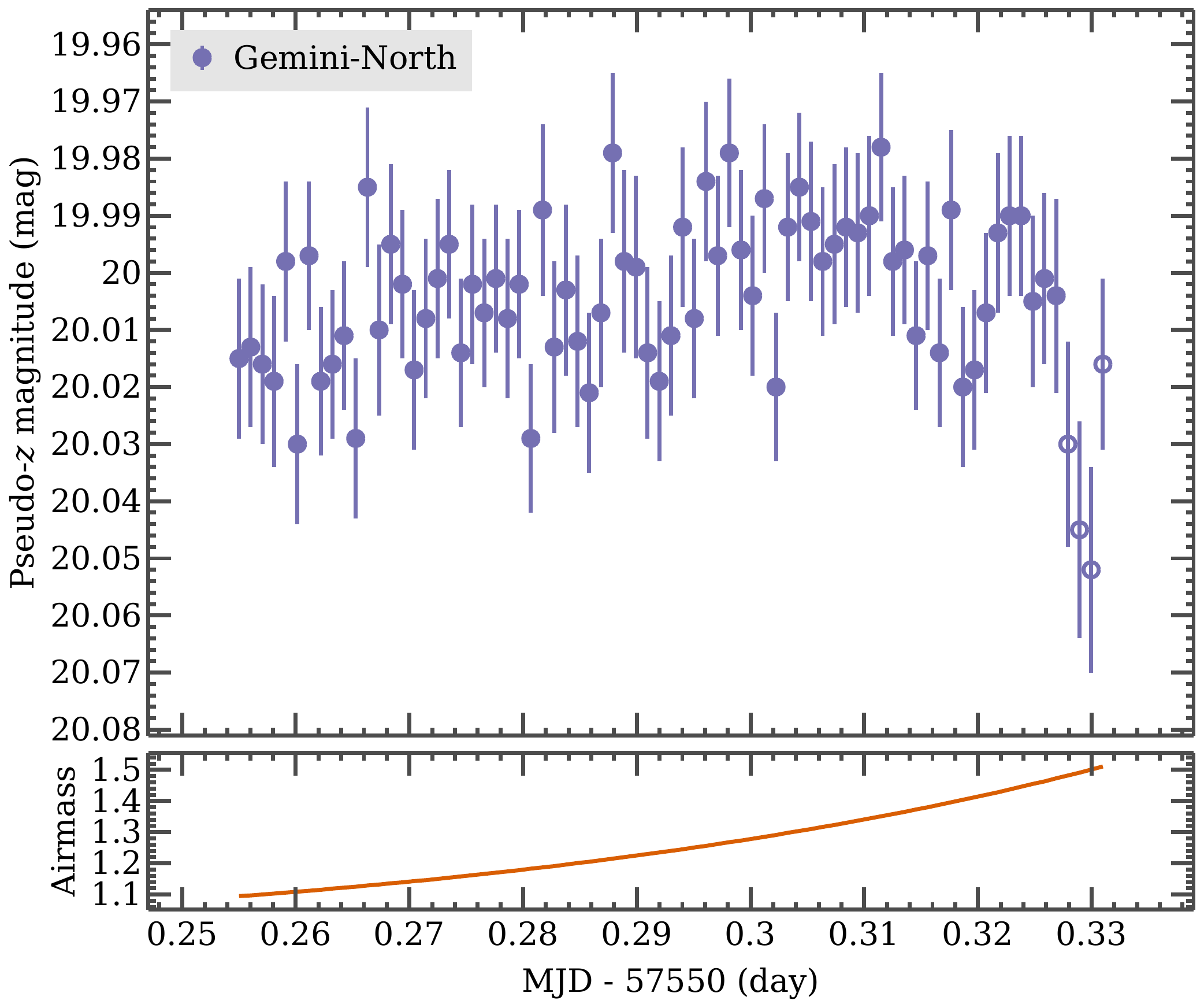}
  \caption{Red optical light curve of \obj{1122+25} from Gemini-North. Un-filled
    circles indicate observations made at high airmass ($z > 1.48$) not
    included in subsequent analysis. The sample cadence is 133~s and the total
    duration of the observation (including the high-airmass points) is
    109~minutes.}
  \label{f.gemini}
\end{figure}

\subsection{Gemini-North}

We observed \obj{1122+25} on UT~date 2016~June~21 with the Gemini Multi-Object
Spectrograph \citep[GMOS-N;][]{hjas+04} on the Gemini-North Telescope (Program
GN-2016A-FT-29) in imaging mode. We used the $z$ (G0304) filter with the e2v
deep depletion detectors. The $z$-band blue edge of 850~nm is set by the
filter, but the red cutoff is set by the detector which extends to
950--1000~nm, approximately 100~nm redder than most CCD detectors.

The data were taken over 1.8~hr and consist of 75 images with 60~s exposure
times. The observations began after the object had already transited the
meridian and spanned an airmass of 1.1 to 1.5. We processed the data in the
standard way using the Gemini IRAF package. We corrected fringing using the
observatory's fringe image and measured aperture photometry for the target
relative to the average of five comparison stars in the field of view to
remove the first-order effects of seeing and airmass variations.
\autoref{f.gemini} shows the resulting light curve. We do not include in our
analysis the final four images made at airmass $z > 1.48$. Calibrating from
the SDSS photometry for these comparison stars yields $z = 20.00$~mag for
\obj{1122+25}. The observed scatter for \obj{1122+25}, ${\pm}0.014$~mag, is
consistent with the expected noise due to photon counting statistics. We rule
out sinusoidal peak-to-peak amplitudes ${\gtrsim}1.5$\% for periods less than
one hour.

\section{Periodicity Analysis}
\label{s.periodicity}

While \autoref{f.gemini} shows no clear evidence for a periodicity in the
Gemini-North data, the peaks in the VLA light curve (\autoref{f.vla}) suggest
a possible periodic signal. We investigated this possibility using a phase
dispersion minimization analysis \citep[PDM;][]{the.pdm}. In this technique,
given a trial period $P$, the data are placed into phase bins and the overall
scatter within each bin is summarized with a statistic denoted $\Theta$. Lower
values of $\Theta$ imply less scatter and therefore a better phasing. Unlike
some other periodicity-finding techniques \citep[e.g., the Lomb-Scargle
  periodogram;][]{l76, s82}, PDM analysis therefore makes no assumptions about
the underlying structure of the potential periodic signal. In this work we
used the implementation of PDM provided by version 0.8.4 of the \textsf{pwkit}
Python toolkit\footnote{\url{https://github.com/pkgw/pwkit/}}.

\begin{figure*}[tbp]
  \includegraphics[width=\linewidth]{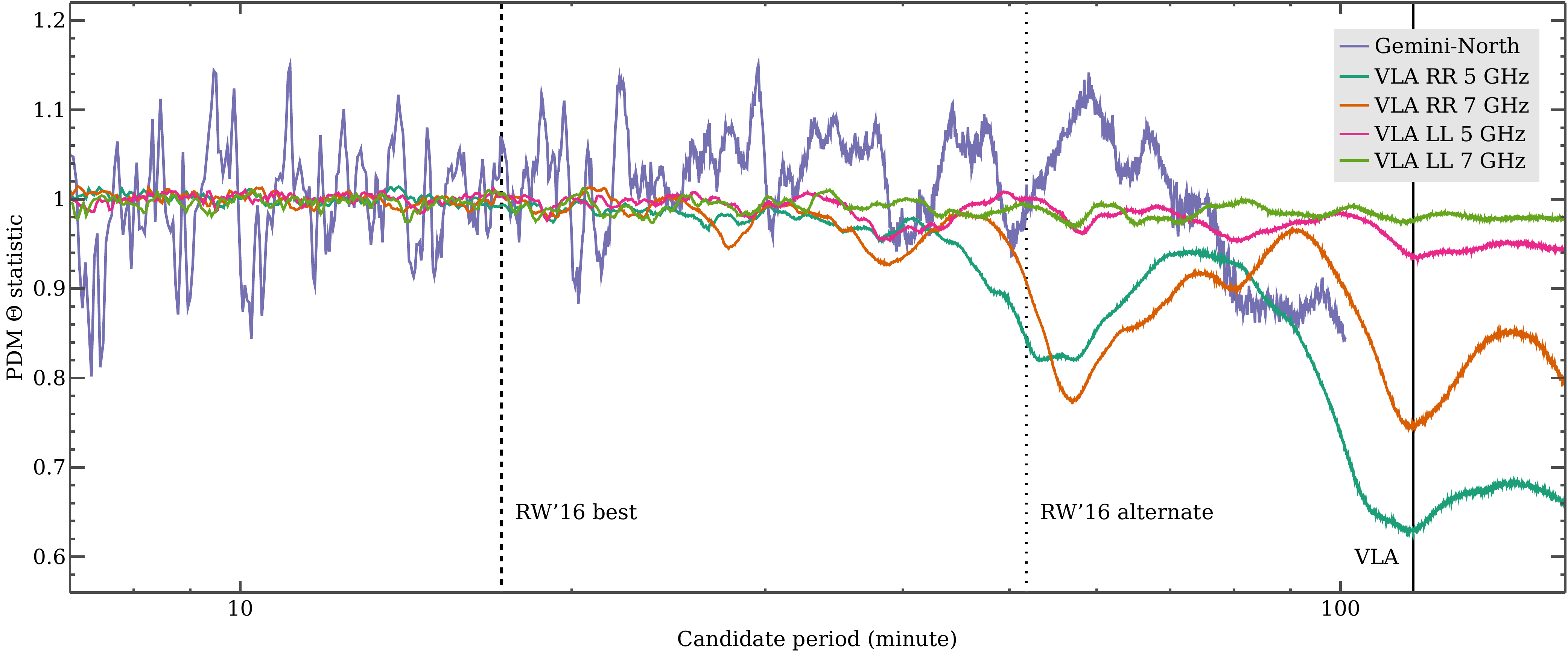}
  \caption{Phase dispersion minimization (PDM) statistic $\Theta$ for trial
    phasings of the VLA and Gemini-North data sets. While there is no clear
    periodicity in the Gemini-North data, the VLA data show multiple minima
    suggesting plausible phasings. \textit{Vertical solid line} shows the best
    phasing of the VLA LL 5~GHz data set, which is also the close to the best
    phasings of the LL 7~GHz and RR 5~GHz light curves. \textit{Vertical
      dashed line} shows the periodicity $P = 17.26$~min suggested by
    \citet{rw16}. Our data do not show evidence for periodic variability at
    this value. \textit{Vertical dotted line} shows another candidate
    periodicity, $P = 51.79$~min, reported by those authors.}
  \label{f.pdm}
\end{figure*}

\begin{figure}[tbp]
  \includegraphics[width=\linewidth]{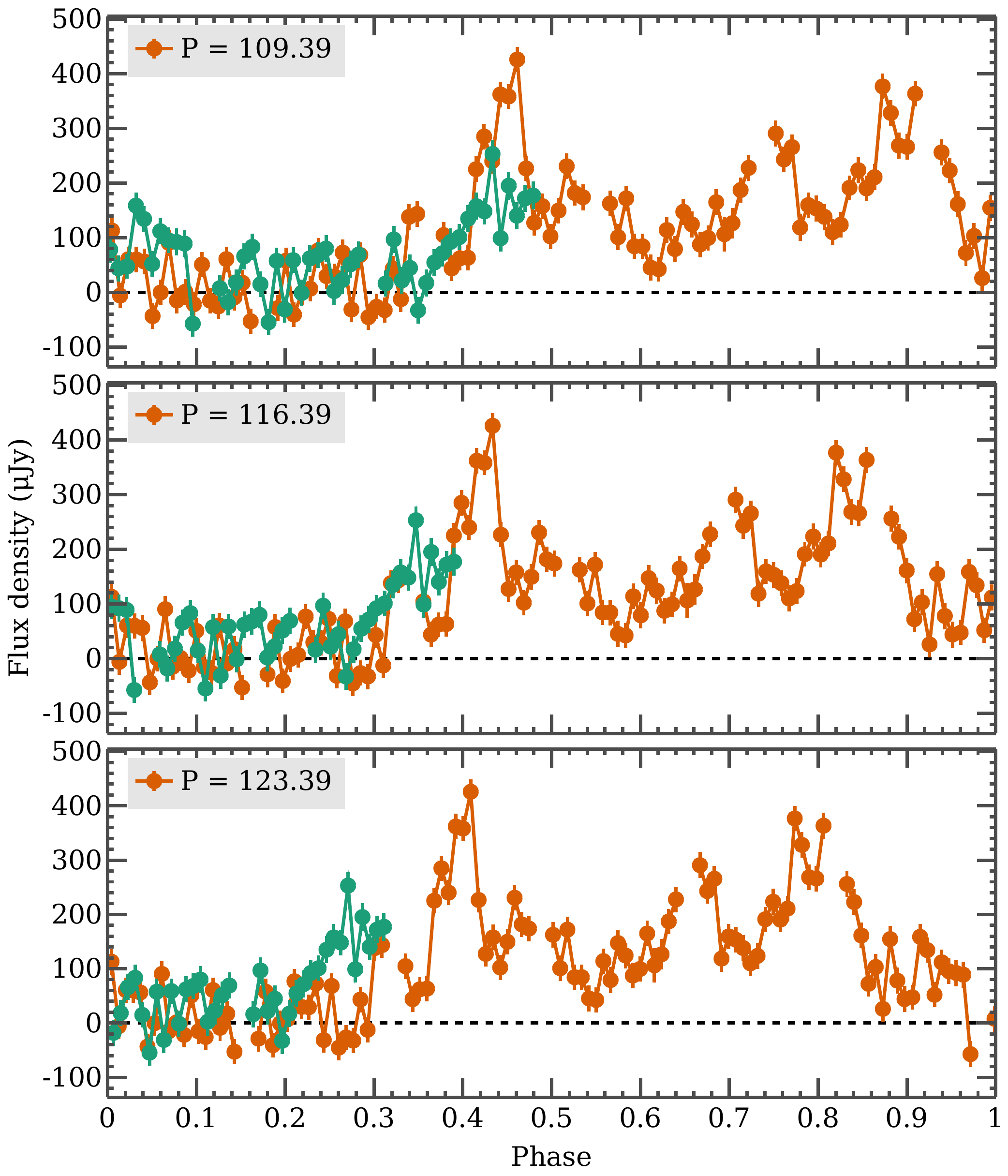}
  \caption{The VLA RR 5~GHz light curve phased at different candidate
    periodicities. Different colors represent different successive rotations.}
  \label{f.phased}
\end{figure}

\autoref{f.pdm} plots $\Theta(P)$ in the VLA and Gemini-North data sets, where
we subdivided the VLA data by polarization and frequency sub-band. The periods
range from one minute to just below the respective durations of the
observations. The PDM analysis does not reveal evidence of a periodic signal
in the Gemini-North data. The VLA RR data, on the other hand, show clear but
broad minima at periods of \apx55 and \apx116~minutes. The LL data show weak
evidence for similar periodicities. Three out of the four VLA $\Theta$ curves
have their global minimum near the larger of these periodicities. We note that
the four curves are derived from data that are statistically independent in
principle. A time-dependent calibration error could induce correlations
between the different polarization and frequency groupings, but PDM analysis
of a nearby reference source shows no indications of such an effect.

Inspection of \autoref{f.vla} shows that the smaller period minimum represents
an approximate phasing of the three peaks seen in the VLA RR 5~GHz data at
$\dt = 50$, $100$, and $160$, while the larger is approximately its
first sub-harmonic, aligning the first and third of these peaks as well as the
troughs of \apx zero RR emission before them. Assuming the shorter (longer)
candidate period, our VLA observation spanned 2.9 (1.4) cycles of emission.
Although the longer candidate period result is therefore associated with
substantially worse phase coverage, with only 0.4 of a period having any
redundant data, the scatter at that phasing is lower than that at the shorter
period, as reflected by the fact that it indeed corresponds to the global
minimum of the PDM $\Theta$ statistic computed for this data set. We proceed
under the assumption that the longer candidate periodicity corresponds to the
rotation period of \obj{1122+25}.

A Monte Carlo assessment of the uncertainty on this periodicty, computed by
adding noise to the observed data and rerunning the PDM analysis, suggests an
uncertainty of 0.7~min. This value, however, understates the true uncertainty
given the limited available data. In \autoref{f.phased} we plot the VLA RR
5~GHz data phased at our preferred periodicity and two periods offset from the
PDM-minimized value by ${\pm}7$~minutes. These phasings are visibly inferior
to the PDM-optimized value but do not appear implausible. Assuming the
emission is indeed periodic, we suggest that a plausible uncertainty on our
measurement is about this this large.

If the Arecibo radio bursts occur at the same periodicity as the VLA
variability, the flare TOAs reported by \citet{rw16} could potentially yield a
more precise constraint on \obj{1122+25}'s periodicity. To investigate this,
we examined possible TOA phasings at periods between 106 and 126 minutes. The
five burst TOAs reported by \citet{rw16} span three primary time scales: the
gap between the first and second TOA is \apx230~days; the gap between the
second and third is \apx18~minutes; and the gaps between the third, fourth,
and fifth TOAs are \apx4~days each. There are therefore reasonable grounds to
argue that the first TOA should not be included in a timing solution, since it
is so distant from the others, and/or that the second or third TOA should not
be included, since their separation is small compared to the VLA light curve
period. Regardless of which, if any, TOAs we discard from the analysis, no
preferred periodicity emerges within the general bounds supported by the VLA
light curve.

\section{Analysis of the Radio Emission}
\label{s.radioprops}

The variability of \obj{1122+25} makes it difficult to assess the brightness
of its quiescent radio emission. The mean flux densities during the troughs of
RR emission, $6 < \dt < 24$ and $121 < \dt < 147$, are almost all consistent
with zero. The exception is the 5~GHz sub-band in the second trough, where the
mean flux density is $33.7 \pm 5.0$~\ujy. Comparing to the LL emission at
these times, this elevated RR emission is likely related to the relatively
bright, low-band LL burst that occurs at this time. The mean 5~GHz LL flux
density in this window is $95 \pm 6$~\ujy, implying a $\approx$50\% LCP flare
of total intensity $\text{I} = \text{LL} + \text{RR} = 130 \pm 8$~\ujy. This
polarization fraction is consistent with the Arecibo events, but the flux
density is an order of magnitude lower \citep{rw16}.

The mean LL flux densities of \obj{1122+25} in the window centered between the
two RR flares, $60 < \dt < 96$, are $18 \pm 5$ and $13 \pm 4$~\ujy\ in the 5
and 7~GHz sub-bands, respectively. We did not obtain the calibrator
observations necessary to calibrate the leakage between the VLA's RR and LL
receivers in our observation, so that the LL correlations will include a small
contribution of RR flux, and vice versa. This is typically a 5\% effect,
whereas the observed LL flux density during this short period represents $(11
\pm 3)$\% of the RR flux density. We therefore cannot exclude that the
measured LL flux density in this time window is entirely due to leakage from
the RR emission.

The data suggest that RR and LL components of the radio emission of
\obj{1122+25} both vary periodically (\autoref{f.pdm}), but 180$\degr$ out of
phase, with their respective maxima including strongly polarized flares
lasting \apx15~minutes and their minima consistent with zero emission.
Therefore, while the mean flux density of \obj{1122+25} in our observation
results in an easy VLA detection ($64 \pm 3$~\ujy\ at 6~GHz with a spectral
index $\alpha = -1.5 \pm 0.3$), this T~dwarf may have \textit{no} quiescent
emission, defining this to mean a non-variable, broadband component.
High-sensitivity radio observations with full polarimetric calibration are
needed to confirm this. The radio flares detected by Arecibo are $\gtrsim$200
times brighter than the quiescent component, taking the brightness of the
latter to be $\lesssim$10~\ujy. This ratio is comparable to the extreme values
found in \obj{1047+21} \citep{wbz13} and the M8 dwarf \obj{1048-39}
\citep{bp05}.

While \obj{1122+25} may not produce ``quiescent'' radio emission as defined
above, its time-averaged radio flux density is relevant to the overall
energetics of the nonthermal particle acceleration processes that power its
radio emission. Assuming a 20\% uncertainty on its estimated 17.7~pc distance,
we compute $\lnura = 10^{13.4\pm0.2}$~erg~s$^{-1}$~Hz$^{-1}$ at 6~GHz. This is
typical of radio-active ultracool dwarfs \citep{wcb14}. Using the polynomial
relationship between spectral type and \lbol\ developed by \citet{frf+15}, we
estimate the bolometric luminosity to be $10^{-4.89 \pm 0.13}$~\lsun, and
therefore that in a time-averaged sense, $\lnura/\lbol = 10^{-15.4 \pm
  0.2}$~Hz$^{-1}$. This is the highest value yet measured for an ultracool
dwarf \citep{wcb14, wb15}, a result driven by the late spectral type of
\obj{1122+25}; as shown by \citet{frf+15}, the bolometric luminosities of
brown dwarfs are believed to decrease rapidly as a function of spectral type
for objects later than \apx T3. That being said, we calculate that both of the
two other late T~dwarfs detected at radio wavelengths, \obj{1047+21} and
\obj{1237+65}, are half as radio-bright as \obj{1122+25} by this metric,
having $\lnura/\lbol \apx 10^{-15.7}$~Hz$^{-1}$ \citep{wb15, khp+16}.

\citet{rw16} interpreted the Arecibo radio bursts as being due to the ECMI,
which results in bursty, highly-polarized emission with brightness
temperatures that can significantly exceed the canonical limit of $10^{12}$~K
associated with the inverse Compton catastrophe \citep{kpt69}. The
non-bursting radio emission from ultracool dwarfs is often argued to instead
originate from gyrosynchrotron processes \citeeg{b02}, which results in
steadier, less polarized emission with typically observed brightness
temperatures of \apx$10^{8\text{--}10}$~K \citeeg{ohbr06}. Assuming the
characteristic length scale of the emitting region to be \rjup, we place a
conservative lower limit on the brightness temperature of the VLA emission to
be $T_b \gtrsim 10^{9.2}$~K, consistent with gyrosynchrotron. \citet{rw16}
obtain a higher limit, $4 \times 10^{11}$~K, for the Arecibo bursts, which are
brighter and more probably originate from a compact emission region. However,
while gyrosynchrotron emission can in principle be highly circularly polarized
\citep{d85}, this requires specialized circumstances that are difficult to
achieve in practice. ECMI or perhaps plasma emission is therefore the more
likely origin of the emission that we observed with the VLA. Observations at
higher frequencies could potentially discriminate between these possibilities
because the spectrum of ECMI emission cuts off strongly at frequencies above
the local electron cyclotron frequency $\nu_c = e B / 2 \pi m_e c \approx 2.8
(B / \text{kG})$~GHz, while gyrosynchrotron emission is spectrally smooth.
Under the ECMI model, we can place a somewhat higher limit on the magnetic
field strength of \obj{1122+25} than \citet{rw16} because our observations
extended to higher frequencies. If the rapid event in the 7~GHz sub-band at
$\dt \sim 43$ is due to ECMI, the magnetic field of \obj{1122+25} must reach
strengths of $B \gtrsim 2.5$~kG.

\begin{figure*}[tb]
  \includegraphics[width=\linewidth]{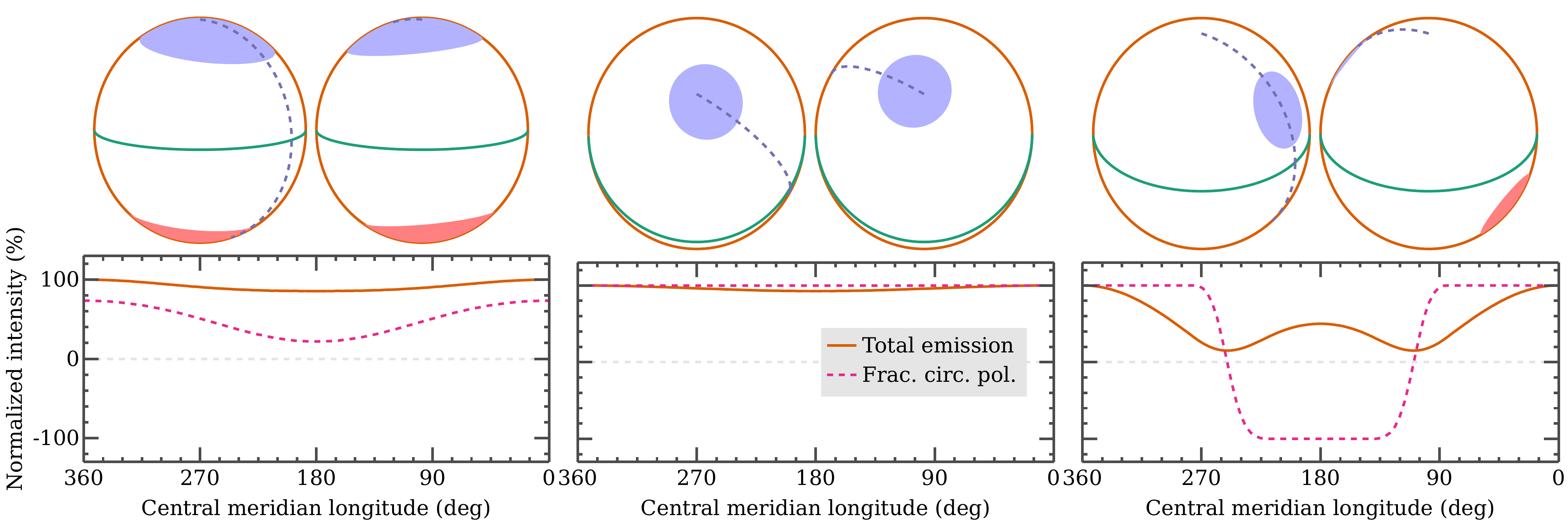}
  \caption{A schematic for interpreting the intensity/polarization light curve
    of \obj{1122+25}. Upper row: orthographic projections of brown dwarf
    configurations. Solid green line is the equator, blue is RR-emitting
    magnetic dipolar cap, red is LL-emitting magnetic dipolar cap, and dashed
    line is zero longitude, defined to intersect both the rotational and
    magnetic poles. Each pair of projections shows the object at central
    meridian longitude $\text{CML} = 300\deg$ and $\text{CML} = 120\deg$. The
    longitude system is planetocentric \citep[see, e.g.,][]{z15}, so that CML
    decreases with time. Lower row: light curves of total intensity
    ($\text{RR} + \text{LL}$; orange solid) and fractional circular
    polarization ($(\text{RR} - \text{LL}) / (\text{RR} + \text{LL})$; purple
    dashed) over one rotation, assuming isotropic emission. Total intensity is
    normalized to its maximum. Left column: configuration with inclination $i
    = 80\deg$, dipole offset $\theta = 6\deg$, cap radius $\phi = 40\deg$.
    Middle column: $i = 20\deg$, $\theta = 6\deg$, $\phi = 20\deg$. Right
    column: $i = 30\deg$, $\theta = 60\deg$, $\phi = 20\deg$. This resembles
    the data in \autoref{f.vla}.}
  \label{f.schematic}
\end{figure*}

\section{Discussion}
\label{s.disc}

We have discovered evidence for variations in the radio emission of
\obj{1122+25} with a tentative period of \apx116~min. The features that drive
this periodicity are the deep troughs in the RR emission, the rising slopes
that follow them, and weaker, low-frequency LL emission features at $\dt \apx
5$ and $130$. Because our VLA observation lasted only 162~min, yielding
overlapped phase coverage of only 0.4 of a rotation, the periodicity we detect
should be regarded as provisional. We also see no compelling signs of
variability at any periodicity in the Gemini-North data, although our
observations did not last as long as this candidate periodicity. Regardless of
the VLA periodicity's validity, however, we do not find evidence for periodic
variability in the radio or red optical emission of \obj{1122+25} at the
period of $17.26313 \pm 0.00001$~min proposed by \citet{rw16}.

If the true rotation period of \obj{1122+25} is 116~min rather than 17~min, it
is not nearly as extreme of an object as initially proposed \citep{rw16}. In
fact, it then becomes a slightly slower rotator than \obj{1047+21}, the only
other T~dwarf to have its rotation period derived from its radio emission
\citep{wb15}, which has $P \approx 106$~min. Assuming a radius of $0.9 \pm
0.15$~\rjup\ for \obj{1122+25}, the same as \obj{1047+21} \citep{vhl+04,
  wb15}, we find an equatorial rotational velocity of \apx60~\kms\ for
\obj{1122+25}.

In these data the radio emission of \obj{1122+25} is strongly circularly
polarized, but the handedness of the polarization changes twice. If this
change in handedness is confirmed to occur periodically, this would be
consistent with a model in which the magnetosphere of \obj{1122+25} is a
dipole tilted with respect to its rotation axis, as proposed by \citet{mbi+11}
to explain VLA observations of the extremely radio-active, benchmark M7~binary
\obj{n33370} \citep[= 2MASS~J13142039$+$1320011; see also][]{sbh+14, wbi+15,
  fdr+16, dfr+16}. We show a schematic of the proposed model in
\autoref{f.schematic}. Each pole produces highly circularly polarized radio
emission, with the oppositely-aligned polar magnetic fields leading to
opposite handedness of the resulting emission. The significantly brighter
emission in the RR polarization mode suggests that the magnetic north pole
(i.e., analogous to Earth's geographic south pole) is more directly pointed
towards Earth \citep{kar+78, kar+78.erratum}. Assuming that the intrinsic
radio emission of each magnetic pole is approximately the same, the shapes of
the observed light curves in each polarization could be used to constrain the
inclination $i$ of the rotation axis of \obj{1122+25} and the angle $\theta$
between this axis and the magnetic dipole axis: for instance, $i = 90\degr$
and $\theta = 0\degr$ leads to an non-variable, unpolarized light curve; while
$i = \theta = 0\degr$ leads to a non-variable, 100\% polarized light curve;
and $i = 0\degr$, $\theta = 90\degr$ leads to out-of-phase curves of 100\%
circular polarization, as observed here. We therefore speculate that the
magnetic dipole axis of \obj{1122+25} may be highly misaligned with its
rotation axis.

The lack of observed $z$-band variability in \obj{1122+25} can only rule out
relatively short (${\lesssim}1$~hr) periods with relatively high
(${\gtrsim}1.5$\%) amplitudes. \citet{mha+15} found that 36\% of T dwarfs vary
in the mid-infrared Spitzer bands at ${>}0.4$\%. \citet{rlja14} argue that at
J band, sinusoidal amplitudes larger than 2\% are found only for L/T
transition objects (spectral types L9--T3.5), but that 80\% of other L/T
dwarfs vary at the 0.5\%--1.6\% level. \citet{hmk15} argued that for T dwarfs,
variability in the red optical (F814W, 700--950~nm) has a higher amplitude
than at J band or Spitzer bands. Given these statistics, \obj{1122+25} cannot
be viewed as an outlier. However, \citet{khp+16} argue for a possible
correlation between radio activity and the presence of optical and infrared
variability which they attribute to aurorae. In this aurora model, the lack of
red-optical variabilty may be more surprising.

\section{Conclusions}
\label{s.conc}

\obj{1122+25} is the fourth T~dwarf with a published detection at radio
wavelengths \citep{rw12, rw16, wbz13, khp+16}. While early theoretical work
motivated a consensus that the magnetic fields of such cool objects should be
negligible \citeeg{ddyr93}, radio surveys have demonstrated that in fact
\apx10\% of these objects host both organized, kG-strength magnetic fields and
the nonthermal particle acceleration processes necessary to drive radio
emission \citep{rw13, rw16, khp+16}. It remains unknown what fraction of the
remaining objects do not possess strong, organized fields at all, and what
fraction possess such fields but not the acceleration processes necessary to
render them detectable via radio emission.

While the bursts from \obj{1122+25} detected by \citet{rw16} are similar to
the bright, highly polarized ECMI bursts detected from a growing list of
ultracool dwarfs, the emission we have detected with the VLA is unique in that
it has both a high duty cycle and strong circular polarization of alternating
handedness. For instance, while the handedness of the circularly polarized
emission from \obj{n33370} also oscillates, it is accompanied by a much more
prominent unpolarized component as well \citep{mbi+11, wbi+15}. We have
speculated that the particular variability pattern we have observed may be due
to a magnetic dipole axis that is highly misaligned with respect to the
object's rotation axis (\autoref{f.schematic}). The lack of unpolarized
emission may also imply that \obj{1122+25} does not possess equatorial belts
of nonthermal plasma as have been hypothesized to surround \obj{n33370} and
other ultracool dwarfs \citep{wcb14, wbi+15}.

Like \obj{1047+21}, \obj{1122+25} was discovered to be a radio emitter with
Arecibo, then followed up with VLA observations \citep{rw12, wbz13}. In both
cases, the interferometric data revealed details of its radio emission that
are inaccessible to Arecibo due to the latter telescope's confusion
limitations and inability to track sources for long ($\gtrsim$3~hr) durations.
Together, these two complementary, world-class observatories are leading the
way in gaining observational insight into the magnetic fields and dynamos of
ultra-cool bodies that are nearing the (exo)planetary mass regime.

\acknowledgments

\textit{Acknowledgments.} The VLA is operated by the National Radio Astronomy
Observatory, a facility of the National Science Foundation operated under
cooperative agreement by Associated Universities, Inc. Based on observations
obtained at the Gemini Observatory processed using the Gemini IRAF package,
which is operated by the Association of Universities for Research in
Astronomy, Inc., under a cooperative agreement with the NSF on behalf of the
Gemini partnership: the National Science Foundation (United States), the
National Research Council (Canada), CONICYT (Chile), Ministerio de Ciencia,
Tecnolog\'{i}a e Innovaci\'{o}n Productiva (Argentina), and Minist\'{e}rio da
Ci\^{e}ncia, Tecnologia e Inova\c{c}\~{a}o (Brazil). This work made use of
NASA's Astrophysics Data System and the SIMBAD database, operated at CDS,
Strasbourg, France.

\facilities{Gemini-North, Karl G. Jansky Very Large Array}

\software{aoflagger, CASA, Gemini IRAF, pwkit}

\bibliography{\jobname}{}

\begin{thebibliography}{}
\providecommand\natexlab[1]{#1}
\providecommand\JournalTitle[1]{#1}

\bibitem[{{Becker} {et~al.}(1995){Becker}, {White}, \& {Helfand}}]{bwh95}
{Becker}, R.~H., {White}, R.~L., \& {Helfand}, D.~J. 1995,
  \href{http://dx.doi.org/10.1086/176166}{\JournalTitle{ApJ}, 450, 559}

\bibitem[{{Berger}(2002)}]{b02}
{Berger}, E. 2002, \href{http://dx.doi.org/10.1086/340301}{\JournalTitle{ApJ},
  572, 503}

\bibitem[{{Berger} {et~al.}(2001){Berger}, {Ball}, {Becker}, {Clarke}, {Frail},
  {Fukuda}, {Hoffman}, {Mellon}, {Momjian}, {Murphy}, {Teng}, {Woodruff},
  {Zauderer}, \& {Zavala}}]{bbb+01}
{Berger}, E., {Ball}, S., {Becker}, K.~M., {et~al.} 2001,
  \href{http://dx.doi.org/10.1038/35066514}{\JournalTitle{Natur}, 410, 338}

\bibitem[{{Berger} {et~al.}(2005){Berger}, {Rutledge}, {Reid}, {Bildsten},
  {Gizis}, {Liebert}, {Mart\'{\i}n}, {Basri}, {Jayawardhana}, {Brandeker},
  {Fleming}, {Johns-Krull}, {Giampapa}, {Hawley}, \& {Schmitt}}]{brr+05}
{Berger}, E., {Rutledge}, R.~E., {Reid}, I.~N., {et~al.} 2005,
  \href{http://dx.doi.org/10.1086/430343}{\JournalTitle{ApJ}, 627, 960}

\bibitem[{{Berger} {et~al.}(2008){Berger}, {Basri}, {Gizis}, {Giampapa},
  {Rutledge}, {Liebert}, {Mart\'{\i}n}, {Fleming}, {Johns-Krull}, {Phan-Bao},
  \& {Sherry}}]{bbg+08}
{Berger}, E., {Basri}, G., {Gizis}, J.~E., {et~al.} 2008,
  \href{http://dx.doi.org/10.1086/529131}{\JournalTitle{ApJ}, 676, 1307}

\bibitem[{{Berger} {et~al.}(2009){Berger}, {Rutledge}, {Phan-Bao}, {Basri},
  {Giampapa}, {Gizis}, {Liebert}, {Mart\'{\i}n}, \& {Fleming}}]{brpb+09}
{Berger}, E., {Rutledge}, R.~E., {Phan-Bao}, N., {et~al.} 2009,
  \href{http://dx.doi.org/10.1088/0004-637x/695/1/310}{\JournalTitle{ApJ}, 695,
  310}

\bibitem[{{Berger} {et~al.}(2010){Berger}, {Basri}, {Fleming}, {Giampapa},
  {Gizis}, {Liebert}, {Mart\'{\i}n}, {Phan-Bao}, \& {Rutledge}}]{bbf+10}
{Berger}, E., {Basri}, G., {Fleming}, T.~A., {et~al.} 2010,
  \href{http://dx.doi.org/10.1088/0004-637x/709/1/332}{\JournalTitle{ApJ}, 709,
  332}

\bibitem[{{Bouvier} {et~al.}(2014){Bouvier}, {Matt}, {Mohanty}, {Scholz},
  {Stassun}, \& {Zanni}}]{bmm+14}
{Bouvier}, J., {Matt}, S.~P., {Mohanty}, S., {et~al.} 2014,
  \href{http://dx.doi.org/10.2458/azu\_uapress\_9780816531240-ch019}{in
  Protostars and Planets VI, ed. H.~{Beuther}, R.~S. {Klessen}, C.~P.
  {Dullemond}, \& T.~{Henning}} (Tucson, AZ, USA: University of Arizona Press),
  433

\bibitem[{{Burgasser} \& {Putman}(2005)}]{bp05}
{Burgasser}, A.~J., \& {Putman}, M.~E. 2005,
  \href{http://dx.doi.org/10.1086/429788}{\JournalTitle{ApJ}, 626, 486}

\bibitem[{{Cook} {et~al.}(2014){Cook}, {Williams}, \& {Berger}}]{cwb14}
{Cook}, B.~A., {Williams}, P. K.~G., \& {Berger}, E. 2014,
  \href{http://dx.doi.org/10.1088/0004-637X/785/1/10}{\JournalTitle{ApJ}, 785,
  10}

\bibitem[{{Cornwell} {et~al.}(2005){Cornwell}, {Golap}, \& {Bhatnagar}}]{cgb05}
{Cornwell}, T.~J., {Golap}, K., \& {Bhatnagar}, S. 2005,
  \href{http://adsabs.harvard.edu/abs/2005ASPC..347...86C}{in Astronomical
  Society of the Pacific Conference Series, Vol. 347, Astronomical Data
  Analysis Software and Systems XIV, ed. P.~{Shopbell}, M.~{Britton}, \&
  R.~{Ebert}}, 86

\bibitem[{{Dulk}(1985)}]{d85}
{Dulk}, G.~A. 1985,
  \href{http://dx.doi.org/10.1146/annurev.aa.23.090185.001125}{\JournalTitle{ARA\&A},
  23, 169}

\bibitem[{{Dupuy} {et~al.}(2016){Dupuy}, {Forbrich}, {Rizzuto}, {Mann},
  {Aller}, {Liu}, {Kraus}, \& {Berger}}]{dfr+16}
{Dupuy}, T.~J., {Forbrich}, J., {Rizzuto}, A., {et~al.} 2016,
  \href{http://arxiv.org/abs/1605.07182}{\JournalTitle{ApJ in press}},
  \href{http://arxiv.org/abs/1605.07182}{{\sffamily arxiv:1605.07182}}

\bibitem[{{Durney} {et~al.}(1993){Durney}, {de Young}, \& {Roxburgh}}]{ddyr93}
{Durney}, B.~R., {de Young}, D.~S., \& {Roxburgh}, I.~W. 1993,
  \href{http://dx.doi.org/10.1007/bf00690652}{\JournalTitle{SoPh}, 145, 207}

\bibitem[{{Filippazzo} {et~al.}(2015){Filippazzo}, {Rice}, {Faherty}, {Cruz},
  {Gordon}, \& {Looper}}]{frf+15}
{Filippazzo}, J.~C., {Rice}, E.~L., {Faherty}, J., {et~al.} 2015,
  \href{http://dx.doi.org/10.1088/0004-637X/810/2/158}{\JournalTitle{ApJ}, 810,
  158}

\bibitem[{{Forbrich} {et~al.}(2016){Forbrich}, {Dupuy}, {Reid}, {Berger},
  {Rizzuto}, {Mann}, {Liu}, {Aller}, \& {Kraus}}]{fdr+16}
{Forbrich}, J., {Dupuy}, T.~J., {Reid}, M.~J., {et~al.} 2016,
  \href{http://arxiv.org/abs/1605.07177}{\JournalTitle{ApJ in press}},
  \href{http://arxiv.org/abs/1605.07177}{{\sffamily arxiv:1605.07177}}

\bibitem[{{Gizis} {et~al.}(2000){Gizis}, {Monet}, {Reid}, {Kirkpatrick},
  {Liebert}, \& {Williams}}]{gmr+00}
{Gizis}, J.~E., {Monet}, D.~G., {Reid}, I.~N., {et~al.} 2000,
  \href{http://dx.doi.org/10.1086/301456}{\JournalTitle{AJ}, 120, 1085}

\bibitem[{{Gizis} {et~al.}(2002){Gizis}, {Reid}, \& {Hawley}}]{grh02}
{Gizis}, J.~E., {Reid}, I.~N., \& {Hawley}, S.~L. 2002,
  \href{http://dx.doi.org/10.1086/340465}{\JournalTitle{AJ}, 123, 3356}

\bibitem[{{G\"{u}del} \& {Benz}(1993)}]{gb93}
{G\"{u}del}, M., \& {Benz}, A.~O. 1993,
  \href{http://dx.doi.org/10.1086/186766}{\JournalTitle{ApJL}, 405, L63}

\bibitem[{{Hallinan} {et~al.}(2006){Hallinan}, {Antonova}, {Doyle}, {Bourke},
  {Brisken}, \& {Golden}}]{had+06}
{Hallinan}, G., {Antonova}, A., {Doyle}, J.~G., {et~al.} 2006,
  \href{http://dx.doi.org/10.1086/508678}{\JournalTitle{ApJ}, 653, 690}

\bibitem[{{Hallinan} {et~al.}(2008){Hallinan}, {Antonova}, {Doyle}, {Bourke},
  {Lane}, \& {Golden}}]{had+08}
---. 2008, \href{http://dx.doi.org/10.1086/590360}{\JournalTitle{ApJ}, 684,
  644}

\bibitem[{{Heinze} {et~al.}(2015){Heinze}, {Metchev}, \& {Kellogg}}]{hmk15}
{Heinze}, A.~N., {Metchev}, S., \& {Kellogg}, K. 2015,
  \href{http://dx.doi.org/10.1088/0004-637X/801/2/104}{\JournalTitle{ApJ}, 801,
  104}

\bibitem[{{Hook} {et~al.}(2004){Hook}, {J\o{}rgensen}, {Allington-Smith},
  {Davies}, {Metcalfe}, {Murowinski}, \& {Crampton}}]{hjas+04}
{Hook}, I.~M., {J\o{}rgensen}, I., {Allington-Smith}, J.~R., {et~al.} 2004,
  \href{http://dx.doi.org/10.1086/383624}{\JournalTitle{PASP}, 116, 425}

\bibitem[{{Kaiser} {et~al.}(1978{\natexlab{a}}){Kaiser}, {Alexander}, {Riddle},
  {Pearce}, \& {Warwick}}]{kar+78}
{Kaiser}, M.~L., {Alexander}, J.~K., {Riddle}, A.~C., {Pearce}, J.~B., \&
  {Warwick}, J.~W. 1978{\natexlab{a}},
  \href{http://dx.doi.org/10.1029/GL005i010p00857}{\JournalTitle{GeoRL}, 5,
  857}

\bibitem[{{Kaiser} {et~al.}(1978{\natexlab{b}}){Kaiser}, {Alexander}, {Riddle},
  {Pearce}, \& {Warwick}}]{kar+78.erratum}
---. 1978{\natexlab{b}},
  \href{http://dx.doi.org/10.1029/GL005i012p01087}{\JournalTitle{Geophysical
  Research Letters}, 5, 1087}

\bibitem[{{Kao} {et~al.}(2016){Kao}, {Hallinan}, {Pineda}, {Escala},
  {Burgasser}, {Bourke}, \& {Stevenson}}]{khp+16}
{Kao}, M.~M., {Hallinan}, G., {Pineda}, J.~S., {et~al.} 2016,
  \href{http://dx.doi.org/10.3847/0004-637X/818/1/24}{\JournalTitle{ApJ}, 818,
  24}

\bibitem[{{Kellermann} \& {Pauliny-Toth}(1969)}]{kpt69}
{Kellermann}, K.~I., \& {Pauliny-Toth}, I. I.~K. 1969,
  \href{http://dx.doi.org/10.1086/180305}{\JournalTitle{ApJL}, 155, 71}

\bibitem[{{Kirkpatrick} {et~al.}(1999){Kirkpatrick}, {Reid}, {Liebert},
  {Cutri}, {Nelson}, {Beichman}, {Dahn}, {Monet}, {Gizis}, \&
  {Skrutskie}}]{krl+99}
{Kirkpatrick}, J.~D., {Reid}, I.~N., {Liebert}, J., {et~al.} 1999,
  \href{http://dx.doi.org/10.1086/307414}{\JournalTitle{ApJ}, 519, 802}

\bibitem[{{Kirkpatrick} {et~al.}(2011){Kirkpatrick}, {Cushing}, {Gelino},
  {Griffith}, {Skrutskie}, {Marsh}, {Wright}, {Mainzer}, {Eisenhardt},
  {McLean}, {Thompson}, {Bauer}, {Benford}, {Bridge}, {Lake}, {Petty},
  {Stanford}, {Tsai}, {Bailey}, {Beichman}, {Bloom}, {Bochanski}, {Burgasser},
  {Capak}, {Cruz}, {Hinz}, {Kartaltepe}, {Knox}, {Manohar}, {Masters},
  {Morales-Calder\'{o}n}, {Prato}, {Rodigas}, {Salvato}, {Schurr}, {Scoville},
  {Simcoe}, {Stapelfeldt}, {Stern}, {Stock}, \& {Vacca}}]{kcg+11}
{Kirkpatrick}, J.~D., {Cushing}, M.~C., {Gelino}, C.~R., {et~al.} 2011,
  \href{http://dx.doi.org/10.1088/0067-0049/197/2/19}{\JournalTitle{ApJS}, 192,
  19}

\bibitem[{{Kirkpatrick} {et~al.}(2012){Kirkpatrick}, {Gelino}, {Cushing},
  {Mace}, {Griffith}, {Skrutskie}, {Marsh}, {Wright}, {Eisenhardt}, {McLean},
  {Mainzer}, {Burgasser}, {Tinney}, {Parker}, \& {Salter}}]{kgc+12}
{Kirkpatrick}, J.~D., {Gelino}, C.~R., {Cushing}, M.~C., {et~al.} 2012,
  \href{http://dx.doi.org/10.1088/0004-637X/753/2/156}{\JournalTitle{ApJ}, 753,
  156}

\bibitem[{{Lomb}(1976)}]{l76}
{Lomb}, N.~R. 1976,
  \href{http://dx.doi.org/10.1007/BF00648343}{\JournalTitle{Ap\&SS}, 39, 447}

\bibitem[{{Luyten}(1926)}]{l26}
{Luyten}, W.~J. 1926,
  \href{http://adsabs.harvard.edu/abs/1926BHarO.835....2L}{\JournalTitle{Harvard
  College Observatory Bulletin}, 835, 2}

\bibitem[{{Mart\'{\i}n} {et~al.}(1999){Mart\'{\i}n}, {Delfosse}, {Basri},
  {Goldman}, {Forveille}, \& {Zapatero Osorio}}]{mdb+99}
{Mart\'{\i}n}, E.~L., {Delfosse}, X., {Basri}, G., {et~al.} 1999,
  \href{http://dx.doi.org/10.1086/301107}{\JournalTitle{AJ}, 118, 2466}

\bibitem[{{McLean} {et~al.}(2011){McLean}, {Berger}, {Irwin}, {Forbrich}, \&
  {Reiners}}]{mbi+11}
{McLean}, M., {Berger}, E., {Irwin}, J., {Forbrich}, J., \& {Reiners}, A. 2011,
  \href{http://dx.doi.org/10.1088/0004-637x/741/1/27}{\JournalTitle{ApJ}, 741,
  27}

\bibitem[{{McMullin} {et~al.}(2007){McMullin}, {Waters}, {Schiebel}, {Young},
  \& {Golap}}]{the.casa}
{McMullin}, J.~P., {Waters}, B., {Schiebel}, D., {Young}, W., \& {Golap}, K.
  2007, \href{http://adsabs.harvard.edu/abs/2007ASPC..376..127M}{in
  Astronomical Society of the Pacific Conference Series, Vol. 376, Astronomical
  Data Analysis Software and Systems XVI, ed. R.~A. {Shaw}, F.~{Hill}, \& D.~J.
  {Bell}}, 127

\bibitem[{{Metchev} {et~al.}(2015){Metchev}, {Heinze}, {Apai}, {Flateau},
  {Radigan}, {Burgasser}, {Marley}, {Artigau}, {Plavchan}, \&
  {Goldman}}]{mha+15}
{Metchev}, S.~A., {Heinze}, A., {Apai}, D., {et~al.} 2015,
  \href{http://dx.doi.org/10.1088/0004-637X/799/2/154}{\JournalTitle{ApJ}, 799,
  154}

\bibitem[{{Offringa} {et~al.}(2010){Offringa}, {de Bruyn}, {Biehl}, {Zaroubi},
  {Bernardi}, \& {Pandey}}]{odbb+10}
{Offringa}, A.~R., {de Bruyn}, A.~G., {Biehl}, M., {et~al.} 2010,
  \href{http://dx.doi.org/10.1111/j.1365-2966.2010.16471.x}{\JournalTitle{MNRAS},
  405, 155}

\bibitem[{{Offringa} {et~al.}(2012){Offringa}, {van de Gronde}, \&
  {Roerdink}}]{ovdgr12}
{Offringa}, A.~R., {van de Gronde}, J.~J., \& {Roerdink}, J. B. T.~M. 2012,
  \href{http://dx.doi.org/10.1051/0004-6361/201118497}{\JournalTitle{A\&A},
  539, A95}

\bibitem[{{Ossendrijver}(2003)}]{o03}
{Ossendrijver}, M. 2003,
  \href{http://dx.doi.org/10.1007/s00159-003-0019-3}{\JournalTitle{A\&ARv}, 11,
  287}

\bibitem[{{Osten} {et~al.}(2006){Osten}, {Hawley}, {Bastian}, \&
  {Reid}}]{ohbr06}
{Osten}, R.~A., {Hawley}, S.~L., {Bastian}, T.~S., \& {Reid}, I.~N. 2006,
  \href{http://dx.doi.org/10.1086/498345}{\JournalTitle{ApJ}, 637, 518}

\bibitem[{{Radigan} {et~al.}(2014){Radigan}, {Lafreni\`{e}re}, {Jayawardhana},
  \& {Artigau}}]{rlja14}
{Radigan}, J., {Lafreni\`{e}re}, D., {Jayawardhana}, R., \& {Artigau}, E. 2014,
  \href{http://dx.doi.org/10.1088/0004-637X/793/2/75}{\JournalTitle{ApJ}, 793,
  75}

\bibitem[{{Route} \& {Wolszczan}(2012)}]{rw12}
{Route}, M., \& {Wolszczan}, A. 2012,
  \href{http://dx.doi.org/10.1088/2041-8205/747/2/l22}{\JournalTitle{ApJL},
  747, L22}

\bibitem[{{Route} \& {Wolszczan}(2013)}]{rw13}
---. 2013,
  \href{http://dx.doi.org/10.1088/0004-637X/773/1/18}{\JournalTitle{ApJ}, 773,
  18}

\bibitem[{{Route} \& {Wolszczan}(2016)}]{rw16}
---. 2016,
  \href{http://dx.doi.org/10.3847/2041-8205/821/2/L21}{\JournalTitle{ApJL},
  821, L21}

\bibitem[{{Saar} \& {Linsky}(1985)}]{sl85}
{Saar}, S.~H., \& {Linsky}, J.~L. 1985,
  \href{http://dx.doi.org/10.1086/184578}{\JournalTitle{ApJ}, 299, 47}

\bibitem[{{Sault} \& {Wieringa}(1994)}]{the.mfs}
{Sault}, R.~J., \& {Wieringa}, M.~H. 1994,
  \href{http://adsabs.harvard.edu/abs/1994A\%26AS..108..585S}{\JournalTitle{A\&AS},
  108, 585}

\bibitem[{{Scargle}(1982)}]{s82}
{Scargle}, J.~D. 1982,
  \href{http://dx.doi.org/10.1086/160554}{\JournalTitle{ApJ}, 263, 835}

\bibitem[{{Schatzman}(1967)}]{s67}
{Schatzman}, E. 1967,
  \href{http://dx.doi.org/10.1007/BF00151366}{\JournalTitle{SoPh}, 1, 411}

\bibitem[{{Schlieder} {et~al.}(2014){Schlieder}, {Bonnefoy}, {Herbst},
  {L\'{e}pine}, {Berger}, {Henning}, {Skemer}, {Chauvin}, {Rice}, {Biller},
  {Girard}, {Lagrange}, {Hinz}, {Defr\`{e}re}, {Bergfors}, {Brandner},
  {Lacour}, {Skrutskie}, \& {Leisenring}}]{sbh+14}
{Schlieder}, J., {Bonnefoy}, M., {Herbst}, T.~M., {et~al.} 2014,
  \href{http://dx.doi.org/10.1088/0004-637x/783/1/27}{\JournalTitle{ApJ}, 783,
  27}

\bibitem[{{Stellingwerf}(1978)}]{the.pdm}
{Stellingwerf}, R.~F. 1978,
  \href{http://dx.doi.org/10.1086/156444}{\JournalTitle{ApJ}, 224, 953}

\bibitem[{{Stelzer} {et~al.}(2006){Stelzer}, {Micela}, {Flaccomio},
  {Neuh\"{a}user}, \& {Jayawardhana}}]{smf+06}
{Stelzer}, B., {Micela}, G., {Flaccomio}, E., {Neuh\"{a}user}, R., \&
  {Jayawardhana}, R. 2006,
  \href{http://dx.doi.org/10.1051/0004-6361:20053677}{\JournalTitle{A\&A}, 448,
  293}

\bibitem[{{Treumann}(2006)}]{t06}
{Treumann}, R. 2006,
  \href{http://dx.doi.org/10.1007/s00159-006-0001-y}{\JournalTitle{A\&ARv}, 13,
  229}

\bibitem[{{van Maanen}(1940)}]{vm40}
{van Maanen}, A. 1940,
  \href{http://dx.doi.org/10.1086/144192}{\JournalTitle{ApJ}, 91, 503}

\bibitem[{{Vrba} {et~al.}(2004){Vrba}, {Henden}, {Luginbuhl}, {Guetter},
  {Munn}, {Canzian}, {Burgasser}, {Davy Kirkpatrick}, {Fan}, {Geballe},
  {Golimowski}, {Knapp}, {Leggett}, {Schneider}, \& {Brinkmann}}]{vhl+04}
{Vrba}, F.~J., {Henden}, A.~A., {Luginbuhl}, C.~B., {et~al.} 2004,
  \href{http://dx.doi.org/10.1086/383554}{\JournalTitle{AJ}, 127, 2948}

\bibitem[{{West} {et~al.}(2008){West}, {Hawley}, {Bochanski}, {Covey}, {Reid},
  {Dhital}, {Hilton}, \& {Masuda}}]{whb+08}
{West}, A.~A., {Hawley}, S.~L., {Bochanski}, J.~J., {et~al.} 2008,
  \href{http://dx.doi.org/10.1088/0004-6256/135/3/785}{\JournalTitle{AJ}, 135,
  785}

\bibitem[{{West} {et~al.}(2004){West}, {Hawley}, {Walkowicz}, {Covey},
  {Silvestri}, {Raymond}, {Harris}, {Munn}, {McGehee}, {Ivezi\'{c}}, \&
  {Brinkmann}}]{whw+04}
{West}, A.~A., {Hawley}, S.~L., {Walkowicz}, L.~M., {et~al.} 2004,
  \href{http://dx.doi.org/10.1086/421364}{\JournalTitle{AJ}, 128, 426}

\bibitem[{{Williams} \& {Berger}(2015)}]{wb15}
{Williams}, P. K.~G., \& {Berger}, E. 2015,
  \href{http://dx.doi.org/10.1088/0004-637X/808/2/189}{\JournalTitle{ApJ}, 808,
  189}

\bibitem[{{Williams} {et~al.}(2015){Williams}, {Berger}, {Irwin},
  {Berta-Thompson}, \& {Charbonneau}}]{wbi+15}
{Williams}, P. K.~G., {Berger}, E., {Irwin}, J., {Berta-Thompson}, Z.~K., \&
  {Charbonneau}, D. 2015,
  \href{http://dx.doi.org/10.1088/0004-637X/799/2/192}{\JournalTitle{ApJ}, 799,
  192}

\bibitem[{{Williams} {et~al.}(2013){Williams}, {Berger}, \& {Zauderer}}]{wbz13}
{Williams}, P. K.~G., {Berger}, E., \& {Zauderer}, B.~A. 2013,
  \href{http://dx.doi.org/10.1088/2041-8205/767/2/l30}{\JournalTitle{ApJL},
  767, L30}

\bibitem[{{Williams} {et~al.}(2014){Williams}, {Cook}, \& {Berger}}]{wcb14}
{Williams}, P. K.~G., {Cook}, B.~A., \& {Berger}, E. 2014,
  \href{http://dx.doi.org/10.1088/0004-637X/785/1/9}{\JournalTitle{ApJ}, 785,
  9}

\bibitem[{{Wu} \& {Lee}(1979)}]{the.ecmi}
{Wu}, C.~S., \& {Lee}, L.~C. 1979,
  \href{http://dx.doi.org/10.1086/157120}{\JournalTitle{ApJ}, 230, 621}

\bibitem[{{Zangari}(2015)}]{z15}
{Zangari}, A. 2015,
  \href{http://dx.doi.org/10.1016/j.icarus.2014.10.040}{\JournalTitle{Icarus},
  245, 93}

\end{thebibliography}

\end{document}